\documentclass[doublecol]{epl2}

\title{Nonextensivity in the solar magnetic activity during
the increasing phase of solar Cycle 23}
\shorttitle{Nonextensivity in the solar magnetic activity}

\author{D. B. de Freitas\thanks{E-mail: \email{danielbrito@dfte.ufrn.br}}
\and J. R. De Medeiros\thanks{E-mail: \email{renan@dfte.ufrn.br}} }
\shortauthor{D. B. de Freitas \etal}

\institute{
  \inst{} Departamento de F\'{\i}sica,
    Universidade Federal do Rio
    Grande do Norte, 59072-970
    Natal,  RN, Brazil\\
}

\abstract{
In this paper we analyze the behavior of the daily Sunspot Number from the
Sunspot Index Data Center (SIDC), the mean Magnetic Field strength from
the National Solar Observatory/Kitt Peak (NSO/KP) and Total Solar
Irradiance means from Virgo/SoHO, in the context of the
$q$--Triplet which emerges
within nonextensive statistical mechanics. Distributions for the
mean solar Magnetic Field show two different behaviors, with a
$q$--Gaussian for scales of 1 to 16 days and a Gaussian for scales longer
than 32 days. The latter corresponds to an equilibrium state.
Distributions for Total Solar Irradiance also show two different
behaviors (approximately Gaussian) for scales of 128 days and longer,
consistent with statistical equilibrium and $q$--Gaussian for scales $<$ 128
days. Distributions for the Sunspot Number show a $q$--Gaussian
independent of timescales, consistent with a nonequilibrium state. The
values obtained
(``$q$--Triplet''$\equiv$$\{$$q$$_{stat}$,$q$$_{sen}$,$q$$_{rel}$$\}$)
demonstrate that the Gaussian or $q$--Gaussian behavior of the
aforementioned data depends significantly on timescales. These results
point to strong multifractal behavior of the dataset analyzed, with the
multifractal level decreasing from Sunspot Number to Total Solar
Irradiance. In addition, we found a numerically satisfied dual
relation between $q_{stat}$ and $q_{sen}$.}

\begin{document}

\maketitle

\section{Introduction}
One of the most important characteristics of solar magnetic activity is
its periodic and quasiperiodic variations at different timescales. These
cyclic phenomena include the sunspot cycle with a mean period of 11 years
and 152--158 day periodicity observed in the occurrence of high-energy
solar flares. Explaining their root-cause has become a puzzling question
in solar physics. However, studies on the cyclic behavior of the Sun's
large scale magnetic field have provided important information about
different solar physical processes. Oliver et al. \cite{oliver1998}
established that 158--day periodicity in sunspots and high-energy solar
flares are related to a periodic emergence of magnetic flux that appears
near the maxima of some solar cycles. More recently, Knaack et al.
\cite{knaack2005} provided evidence for a variety of quasi-periodic
oscillations in the solar photospheric magnetic field ranging from 24 days
to 22 years.

\begin{figure*}
\centering
\resizebox{0.90\textwidth}{!}{%
  \includegraphics{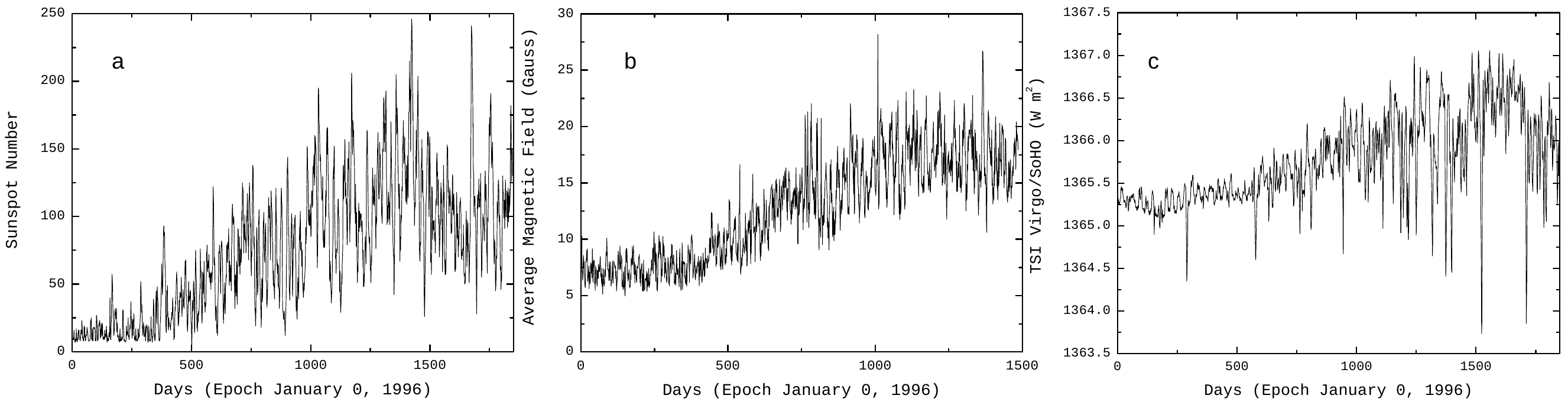}
  }
\caption{Time series of daily values for (a) Sunspot Number computed from
the SIDC database,
(b) Magnetic Field computed from NSO/KP and for (c) Total
Solar Irradiance as
observed by Virgo/SoHO.}
\label{fig0}
\end{figure*}

The statistical properties of solar magnetic activity have been
investigated by different authors in the context of chaos theory
\cite{veronig2000}, multifractal analysis \cite{abramenko2005}, and
nonextensive statistical mechanics \cite{burlaga2004,burlaga2009}. The
Tsallis' nonextensive theory is a generalization of Boltzmann-Gibbs
statistical mechanics (B-G statistics) for out of thermal equilibrium
systems, described by the \textsl{entropic index} $q$ ($q=1$ in the B--G
statistics; for further details: \cite{tsallis1988,abe2000,gell2004}). It
has been successfully applied to many complex physical systems as solar
magnetic activity \cite{burlaga2004,burlaga2009}, including different
astrophysical data presenting nonlinear and nonstationary behavior and
nonequilibrium states. Tsallis \cite{tsallis2004} recently proposed the
existence of a three-index set ({\it q}$_{stat}$,{\it q}$_{sen}$,{\it
q}$_{rel}$), also known as $q$-Triplet, characterized by metastable states
in nonequilibrium, where $q_{stat}>$1, $q_{sen}<$1 and
$q_{rel}>$1. When ({\it q}$_{stat}$,{\it q}$_{sen}$,{\it
q}$_{rel}$)=(1,1,1), the set denotes the B--G thermal equilibrium state.
Burlaga and Vi{\~n}as \cite{burlaga2004} used a
generalized Tsallis probability distribution function to describe the
multiscale structure of fluctuations in solar wind speed and the magnetic
field during the declining phase of solar cycle 23. More recently, these
authors
presented the first physical corroboration of the $q$--Triplet, from the
analyses of the behavior of two sets of daily magnetic field strength
observed by Voyager 1 in the solar wind in 1989 and 2002
\cite{burlaga2005}.

\begin{figure*}
\centering
\resizebox{0.90\textwidth}{!}{%
  \includegraphics{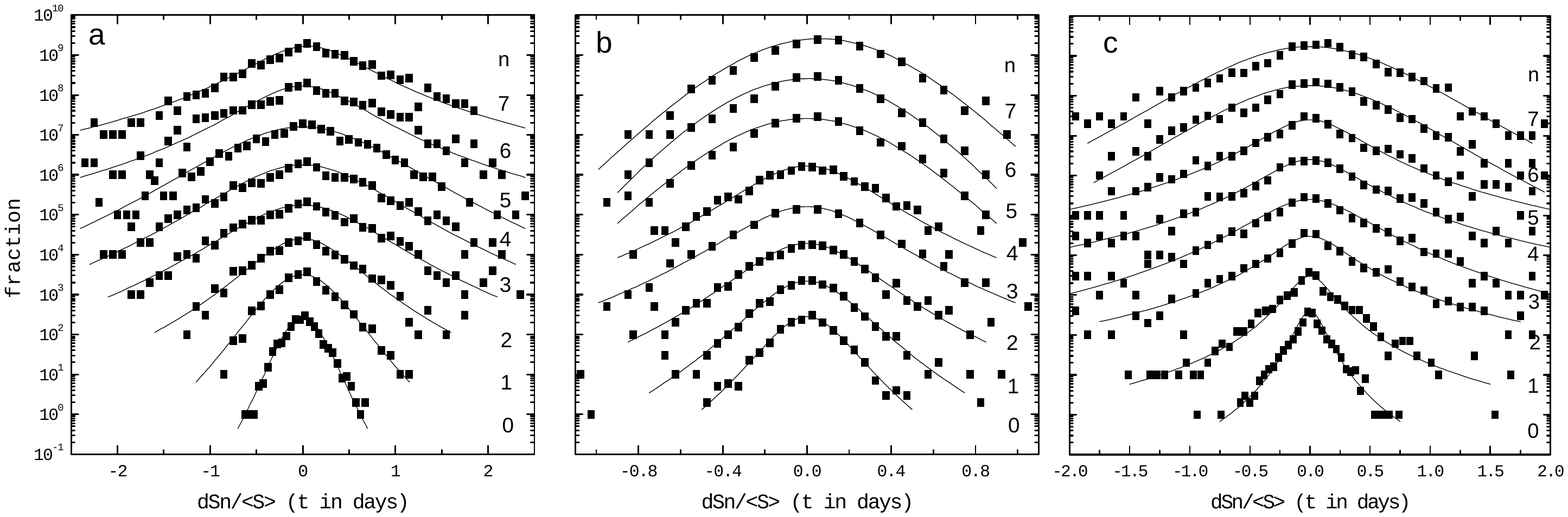}
}
\caption{Distributions of increments for (a) SIDC, (b) NSO/KP and
(c) Virgo/SoHO data. The PDFs are shifted up by a factor of 10 each,
for the sake clarify. The inset shows the timescale $\tau$=2$^{n}$ (days),
from 1 to 128 days for three datasets}, where $n$=0,1,2,...,7.
\label{fig1}
\end{figure*}

\textcolor{black}{The magnetic activity on the solar atmosphere is by far more
complex than such a simplified description, and a variety of dynamical
phenomena are expected, such as fluctuations of the magnetic field and
variability of the solar indices \cite{oliviero1999}. This solar magnetic
activity is a driven nonlinear non-equilibrium system. These properties
can represent a hierarchical structure in phase space, in contrast to the
expected behavior for B--G statistical mechanics (i.e., uniformly occupied
phase space). }

The main aim of this study is to perform an analysis of  the 
behavior of physical
parameters directly reflecting the solar magnetic  activity in the 
context Tsallis $q$--Triplet's
formalism, and to compare the properties of this $q$--Triplet with those 
expected for a metastable or
quasi-stationary dynamical system described by nonextensive statistics. The
physical parameters analyzed are the solar indices: (i) Sunspot Number, 
(ii) Magnetic Field
strength and (iii) Total Solar Irradiance, measured from observations
carried out between 1996 and 2001. These three datasets offer a unique
possibility to analyze the relationship between the number of
solar indices (e.g., sunspots and magnetic flux) and the nonequilibrium
level, the degree of complexity, as well as the relaxation
process, in solar activity diagnostics. The present analysis offers 
also a possibility
for a check of the validity of $q$--Central Limit Theorem, the so-called 
$q$--CLT, recently
conjectured by Umarov, Tsallis and Gell--Mann \cite{umarov2008}, in the 
context of
physical sub--systems as those associated to the solar magnetic activity.

\section{Data and Methods}
Three diagnostics of solar magnetic activity were selected for this study:
(i) daily international Sunspot Number (SN) from the Sunspot Index Data
Center (SIDC); (ii) daily full-disk
measurements of mean line-of-sight component of unsigned Magnetic Field
strength (MF) from the National Solar Observatory/Kitt Peak
(NSO/KP). These data do not cover the
overall period ($\sim$20\% are missing). These temporal gaps are
reconstructed by linear interpolation of the daily values; and, (iii)
daily means of Total Solar Irradiance (TSI) from Virgo/SoHO data.
Observations cover the period
January 1996 \textcolor{black}{(cycle minimum)} to September 2001 \textcolor{black}{(cycle
maximum)}. This time interval corresponds to the increasing phase of cycle
23, \textcolor{black}{which reached its maximum around 2001, thereafter declining
until December 2007. At present, the Sun is known to be in solar activity
Cycle 24}. According \textcolor{black}{to} De Toma et al. \cite{detoma2004} cycle 23
is magnetically weaker and simpler than its immediate predecessors, 21 and
22. \textcolor{black}{Figure \ref{fig0} shows the variability in three solar
activity indices in this phase of cycle 23, SN and MF (without
interpolation) and TSI}.

\textcolor{black}{Virgo/SoHO data reveal the intrinsic variability of the Sun:
from a few seconds up to the 11-year cycle. Lanza et al. \cite{lanza2003}
pointed out that different contributions affect variations of TSI: (i)
shorter timescales ($\leq$2 days), associated to active regions (solar
micro-variability); (ii) longer timescales associated to active region
evolution up to about 30 days (rotational modulation period), up to
periodicities of $\sim$200 days; (iii) timescales longer than 200 days;
and, (iv) very long timescales associated to an 11--year activity cycle.
We are interested in scales between 1 and 200 days. NSO/KP data are
average values over the full solar disk, where several solar indices can
contribute to its variation, such as, sunspots, faculae or plages, whereas
SIDC data are a measure of the general state of solar magnetic
activity. According to De Toma et al. \cite{detoma2004}, TSI variability
during the increasing phase of cycle 23 may be due to magnetic flux
emergence in these indices.}

\subsection{Determination of the $q$--Triplet}
For time series $S(t)$, fluctuations of increments due to its variability
over timescale $\tau$ is given as $dSn(t)=S(t+\tau)-S(t)$, where $\tau$ is
the scale parameter defined as $2^{n}$ that determines the scale of
fluctuations represented by $dSn(t)$ and $n =$0, 1, 2, ... and so on. This
method has been used to describe magnetic field fluctuations due to solar
wind on a large scale range \cite{burlaga2005,burlaga2009}.

Burlaga and Vi{\~n}as \cite{burlaga2004} \textcolor{black}{pointed out} two factors
to consider when analyzing a complex multiscale system such as the solar
magnetic activity on from 1 hour to 1 year: (i) mechanical properties that
differ at different scales and (ii) the statistical probability structures
of fluctuation components in the series at the several scales. A natural
method for analysis of these structures is statistical mechanics. Two main
procedures are mentioned in the literature. One of these is the well-known
B--G statistical mechanics and the other Tsallis statistics. For systems
studied using nonextensive statistical mechanics according to Tsallis
statistics,  energy probability density function, sensitivity to the
initial conditions and relaxation are described by three entropic indices
$q_{stat}$, $q_{sen}$, and $q_{rel}$. These are referred to as the
$q$-Triplet \cite{tsallis2004,burlaga2005}. \textcolor{black}{The present study
applies the same procedure used by Burlaga and Vi{\~n}as
\cite{burlaga2005} to compute the $q$--Triplet from Voyager 1 data}.

The values of $q_{stat}$ are derived from Probability Distribution
Functions (PDFs), defined as
\begin{equation}
\label{1pdf}
PDF[dSn(t)]=A_{q}[1+(q-1)\beta_{q}(dSn)^{2}]^{\frac{1}{1-q}},
\end{equation}
where the coefficients $A_{q}$, $\beta_{q}$ denote the normalization
constants and $q$ the functions of scale $\tau$. This {\it entropic index}
$q$ is related to the size of the tail in the distributions
\cite{burlaga2009}. Our study uses the Levenberg--Marquardt method
\cite{levenberg1944,marquardt1963} to compute the PDFs with symmetric
Tsallis distribution from Equation (\ref{1pdf}).

\textcolor{black}{The values of $q_{sen}$ can be obtained from the multifractal (or
singularity) spectrum $f(\alpha)$, where $\alpha$ is the singularity
strength or H\"{o}lder exponent.  Spectrum $f(\alpha)$ is derived via a
modified Legendre transform, through the application of the MF-DFA5 method
\cite{kantel2002}. This method consists of a multifractal characterization
of nonstationary time series, based on a generalization of the detrended
fluctuation analysis (DFA). The $q_{sen}$--index denotes sensitivity at
initial conditions. For the present purposes, we used the expression defined
by Lyra and Tsallis \cite{lyra1998} for the relation between $q_{sen}$ and
multifractality in dissipative systems:}

\begin{equation}
\label{1sen}
1/(1-q_{sen})=1/\alpha_{min}-1/\alpha_{max},
\end{equation}
where \textcolor{black}{$\alpha_{min}$ and $\alpha_{max}$} denotes the roots of the
best-fit.

The value of $q_{rel}$, which describes a relaxation process, can be
computed from a scale-dependent correlation coefficient defined by
\begin{equation}
\label{1rel}
C(\tau)=\frac{\left\langle [S(t_{i}+\tau)-\left\langle
S(t_{i})\right\rangle][S(t_{i})-\left\langle
S(t_{i})\right\rangle]\right\rangle}{\left\langle [S(t_{i})-\left\langle
S(t_{i})\right\rangle]^{2}\right\rangle},
\end{equation}
and for Tsallis statistics
\begin{equation}
\label{3rel}
\log C(\tau)=a+s\log\tau,
\end{equation}
where the slope $s$ is given by
\begin{equation}
\label{2rel}
s=1/(1-q_{rel}).
\end{equation}

\begin{figure}
\centering
\resizebox{0.45\textwidth}{!}{%
  \includegraphics{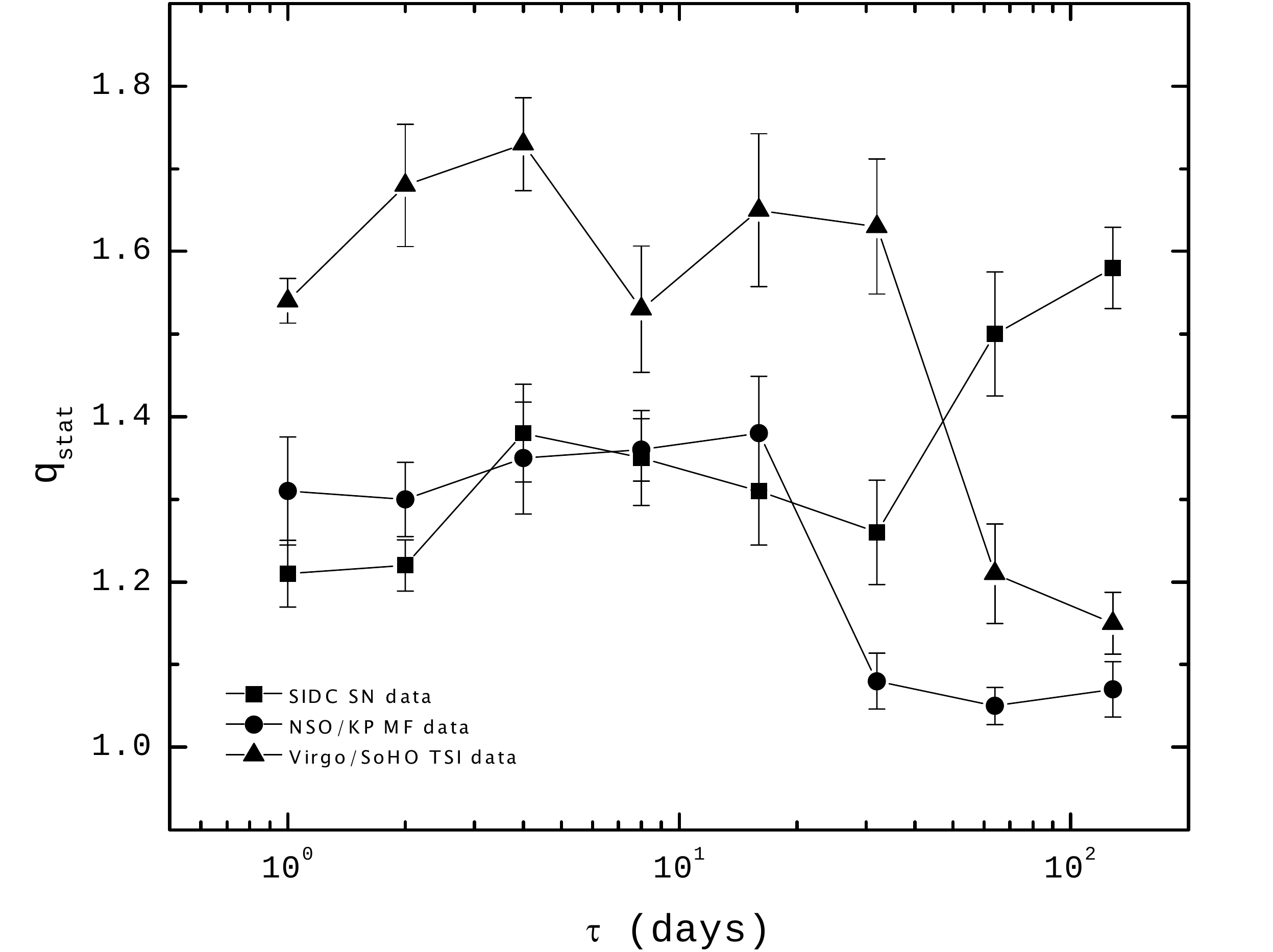}
}
\caption{Evolution of the entropic index $q_{stat}$ as a function of scale
$\tau$, derived from fits of the PDFs in Figures \ref{fig1}\textcolor{black}{(a)},
\ref{fig1}\textcolor{black}{(b)}
and \ref{fig1}\textcolor{black}{(c)} to the $q$--Gaussian distribution. Where
$q$$\sim$1,
consistent with the Gaussian distribution.}
\label{fig1a2}
\end{figure}

\section{Results}
\textcolor{black}{Our results are presented in three subsections, each one
associated to the properties of one of the $q$'s.}

\begin{figure*}
\centering
\resizebox{0.90\textwidth}{!}{%
  \includegraphics{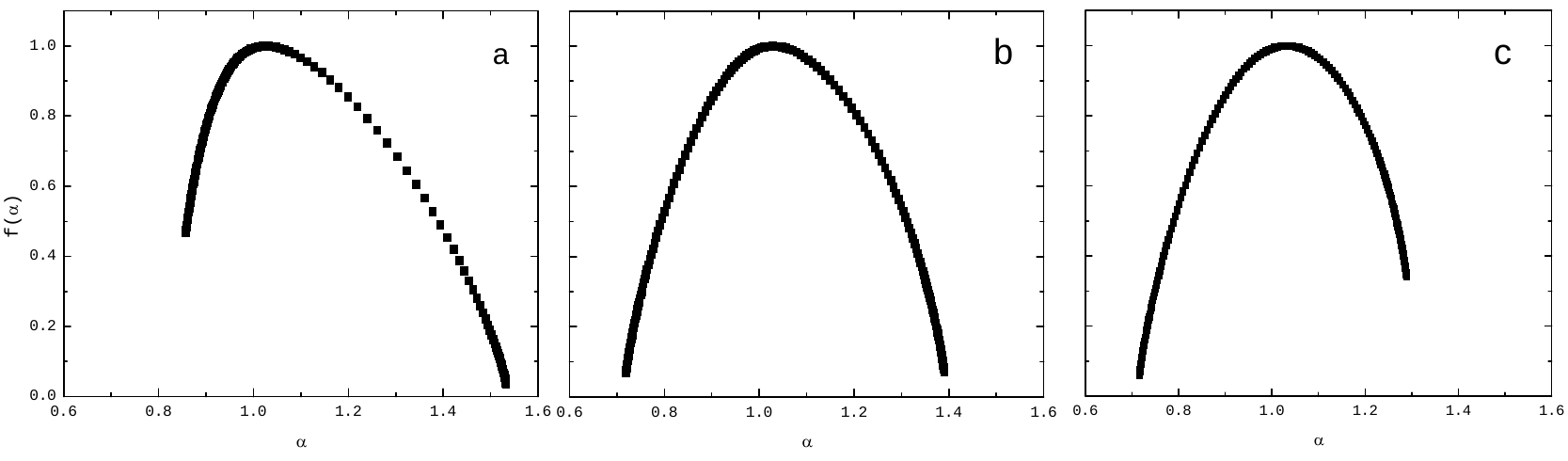}
}
\caption{The symbols are based on measurements of the multifractal
spectrum $f$($\alpha$) versus $\alpha$: (a) SIDC, (b) NSO/KP and (c)
Virgo/SoHO data. All data with 1 day increment.}
\label{fig2}
\end{figure*}

\subsection{\textcolor{black}{On the behavior of Probability Distribution
Functions as a
function of scale: the $q_{stat}$--index}}
PDFs of the signal $S(t)$ were calculated to quantitatively characterize
the stochastic process observed. The signal $S(t)$ represents the
\textcolor{black}{SN, MF and TSI  data for a timescale $\tau$ ranging from 1 to
128 days}.

\textcolor{black}{PDFs} for the three datasets are shown in Figure \ref{fig1}. The
behavior
of the PDFs entropic index changes as a function of scale is clear.
Distributions for the SN shown in Figure \ref{fig1}\textcolor{black}{(a)}
demonstrate $q$--Gaussian behavior independent of timescales, which
corresponds to a nonequlibrium state. However, for the MF the
distributions in Figure \ref{fig1}\textcolor{black}{(b)} show two different
behaviors. A $q$--Gaussian is observed for scales of 1 to 16 days and a
Gaussian for scales larger than 32 days. The latter corresponding to an
equilibrium
state. In relation to the TSI shown in Figure \ref{fig1}\textcolor{black}{(c)},
distributions are approximately Gaussian at scales of \textcolor{black}{128 days}
and higher, consistent with statistical equilibrium, but are non-Gaussian
with long tails and inflection points on scales \textcolor{black}{$<$ 128 days}.

Differences between the PDFs of the dataset analyzed can also be
quantitatively expressed from the nonextensivity $q_{stat}$--index values.
Figure \ref{fig1a2} illustrates the behavior of the $q_{stat}$--index
distribution as a timescale function, derived from fits of the PDFs in
Figure \ref{fig1}. In the NSO/KP data, for timescales less than
approximately 16 days the value for $q_{stat}$$\sim$1.3 is relatively
constant. However, in scales larger than around 32 days,
$q_{stat}$$\sim$1. This is consistent with Gaussian behavior and a lack of
correlation.
\textcolor{black}{For the Virgo/SoHO data, $q_{stat}$$\sim$1.65, at the average, for
scales shorter than about 32 days. This is consistent with the value of
$q$$\sim$1.75
expected for nonlinear systems, where the random variable is a sum of
strongly correlated contributions
\cite{burlaga2005,burlaga2009,tirnakli2007}. In this context,
nonextensivity is consistent in the following aspect: Virgo/SoHO data
is an overlap of several solar indices, such as magnetic activity (flares,
sunspots, faculae and plages) and rotational modulation
\cite{harvey1993,andersen1994,rabello1997,lanza2003}. At the largest scales
$q_{stat}$ decreases and reaches a minimum value of $\sim$1, which is also
consistent with Gaussian behavior and lack of correlation.}

\textcolor{black}{In contrast to the data from NSO/KP and Virgo/SoHO, the
$q_{stat}$--index from SIDC data shows a relatively anomalous behavior,
with $q$$\sim$1.3 at scales shorter than about 32 days and  increasing
 to maximum value of $\sim$1.6 at 128 days. In this case, we found that SN
shows an out
of equilibrium behavior at timescales larger than about 128 days.
As mentioned by Kiyono et al. \cite{kiyono2006},
due to the central limit theorem, Gaussian PDFs at sufficiently long
scales are generally expected, owing to the accumulation of uncorrelated
variations. Nonetheless, Tsallis \cite{tsallis2004} pointed out that
during the metastable or quasi-stationary state, the timescales of an
equilibrium stationary state can be reached, in many cases, infinitely
later.}

\subsection{On the behavior of multifractal spectra $f(\alpha)$: the
$q_{sen}$--index}
\textcolor{black}{The multifractal spectra $f(\alpha)$ were obtained following a
procedure based on the MF-DFA5 method, as mentioned in the Section
\textcolor{black}{Data and Methods}}. The multifractal characterization of these
data is shown in Figure \ref{fig2}. These spectra $f(\alpha)$ calculated
for datasets shows a wide H\"{o}lder exponent interval, with
$\alpha_{mim}$=0.81$\pm$0.08 and $\alpha_{max}$=1.54$\pm$0.04 for the SIDC
data, $\alpha_{mim}$ = 0.71$\pm$0.04 and $\alpha_{max}$ = 1.40$\pm$0.04
for the NSO/KP data and \textcolor{black}{$\alpha_{mim}$ = 0.71$\pm$0.03 and
$\alpha_{max}$ = 1.33$\pm$0.07 for Virgo/SoHO data}. \textcolor{black}{For the
broadness of the multifractal spectrum}
$\Delta\alpha$=$\alpha_{max}-\alpha_{min}$, we obtained 0.73, 0.69 and
0.62 for the daily SN, MF and TSI, respectively.
\textcolor{black}{This behavior denotes the range of fractal exponent present in
the time series and thus gives a measure of the degree of multifractality
of complexity \cite{sen2007}. These results point to a strong multifractal
behavior with the multifractal level decreasing from SIDC to Virgo/SoHO
data.}

\textcolor{black}{Additional parameters can be derived from the multifractal
spectrum,
providing information about the fractal properties as, for exemple, an
asymmetry in the shape of the $f(\alpha)$. Different authors (e.g.:
\cite{telesca2005}) pointed out that a left or right-skewed spectrum
implies high or low fractal exponents, respectively. In this context, the
asymmetry reveals a relative abundance of large or small fluctuations in
the data. Figures \ref{fig2}(a) and (b) show that the multifractal spectra
are right-skewed and slightly right-skewed, respectively, indicating a
dominance of low fractal exponents and a relative abundance of small
fluctuations in SIDC and NSO/KP data. On the other hand, in Figure
\ref{fig2}(c), the multifractal spectrum is slightly left-skewed,
indicating a dominance of high fractal exponents and a relative abundance
of large fluctuations in Virgo/SoHO data. Therefore, in this phase of
solar cycle 23, the relative abundance of small fluctuations is more
predominant in SIDC data than in NSO/KP and Virgo/SoHO data.}

\textcolor{black}{Values of $q_{sen}$--index are directly related to the instability
of the system and to the growth of the entropy. This index was calculated
using Equation (2) based on the values of $\alpha$ in the first paragraph of
the present subsection. The results are
 $-0.71\pm0.10$, $-0.44\pm0.07$ and $-0.52\pm0.10$ for the SN,
MF and TSI, respectively. The values of $q_{sen}<$ 1 denote that its
distribution exhibits weak chaos (i.e., zero Lyapunov exponents, but
\textsl{not} integrable \cite{tsallis2004}) in the full dynamical space of
the system \cite{tsallis2004,burlaga2005}. }

\subsection{On the behavior of scale-dependent correlation: the
$q_{rel}$--index}

\textcolor{black}{The value of $q_{rel}$--index can be determined from the
scale-dependent correlation coefficient $C(\tau)$ according to the
expression given by Equation (\ref{1rel}). In the nonextensive theory this
coefficient should decay following a power law given by Equation
(\ref{3rel}), with increasing $\tau$, where the slope $s$ is given by
Equation
(\ref{2rel}). In general, one expects $C(\tau)$ to evolve with the number of
solar indices on the data and to be a function of the solar cycle. The
values of $q_{rel}$ obtained for SIDC, NSO/KP and Virgo/SoHO
data are 6.9$\pm$4.81 ($R^{2}=0.96$), 12.11$\pm$3.98 ($R^{2}=0.97$) and
7.7$\pm$2.63 ($R^{2}=0.85$), respectively. The $q_{rel}$ values obtained 
show
that the present set of data is not consistent with a power law, owing to
high $q_{rel}$ values, particularly in the NSO/KP data.} Moyano
\cite{moyano2006} suggests that the above procedure to calculate $q_{rel}$
should only be used to describe stochastic processes with linear
correlations. In
other words, the correlation coefficient $C(\tau)$ is not a good
alternative to conveniently describe the non-linearity of a sample.

In B--G statistics, in contrast to the nonextensive theory, the
coefficient $C(\tau)$ should decrease exponentially with an increasing
$\tau$, following a $C(\tau)\propto$ exp$(-\tau/\tau_{c})$ relation, with
$\tau_{c}$ corresponding to the correlation or relaxation time.
\textcolor{black}{Considering our data as consistent with this exponential behavior,
we found, for the parameter $\tau_{c}$, 3.39$\pm$0.05 days ($R^{2}=0.90$),
5.05$\pm$0.38 days ($R^{2}=0.98$) and
3.18$\pm$0.37 days ($R^{2}=0.97$), respectively for the three datasets}.
A correlation time $\tau_{c}$ between 3 and 5 days for \textcolor{black}{these
datasets} corresponds approximately to the half timescale of the
relaxation process obtained by Nandy et al. \cite{nandy2003} for the Sun,
in the magnetic fields of flare-productive solar active regions.
\textcolor{black}{So, if the equilibrium hypothesis of the Sun is considered valid,
then it makes sense to assume $q_{rel}=1$ for the three sets of data. But,
is the Sun in thermal equilibrium? As mentioned by Shibahashi et al.
\cite{shiba1995}, if the Sun is not in thermal equilibrium it takes about
10$^{7}$ years for it to recover its equilibrium state. Therefore, the
condition of thermal equilibrium is a good approximation for $q_{rel}$.}

\section{\textcolor{black}{Discussion and} Conclusions}
We have carried out \textcolor{black}{a new approach} on the nonextensivity
properties of solar magnetic activity from 1996--2001. The study is based
on daily measurements
of Sunspot Numbers from the the Sunspot Index Data Center, daily full-disk
measurements of mean line-of-sight component of unsigned Magnetic Field
strength, from the National Solar Observatory/Kitt Peak, and daily means
of Total Solar Irradiance  from Virgo/SoHO.

\textcolor{black}{The PDFs were calculated for the  three datasets and the
obtained results
show that the entropic indices change as a function of scale. The
distributions for the Magnetic Field strength
show two different behaviors, with a $q$--Gaussian for scales of 1 to 16
days and a Gaussian for scales larger than 32 days. The latter is in
equilibrium state. The Total Solar Irradiance also show two
different behaviors, namely, approximately Gaussian at scales of 128 
days and
larger, consistent with statistical equilibrium and $q$--Gaussian for scales
$<$ 128 days}. In contrast, the distributions for the Sunspot Number show a
$q$--Gaussian independent of timescales and consistent with a
nonequilibrium state for all the timescales. These results also confirm
that the Gaussian or $q$--Gaussian behavior of the data depends
significantly on the solar activity indices.

\textcolor{black}{Our results reveal that during the increasing phase of solar cycle
23, the multifractal spectra of the analyzed datasets differ
significantly. In
particular, we found that  SIDC and NSO/KP  data have
higher complexity than  Virgo/SoHO data. In contrast, Virgo/SoHO
data present a dominance of higher fractal exponents and a relative
abundance of larger fluctuations than for SIDC and NSO/KP data.}

\textcolor{black}{The results above suggest that SIDC
data are out of equilibrium more than those from NSO/KP and Virgo/SoHO.
Again, the multifractal spectra of SIDC are more asymmetric, more weakly
chaotic
(sensitivity to the initial conditions) and also of higher
complexity than the NSO/KP and Virgo/SoHO data. In fact, according to De
Toma et al.
\cite{detoma2004}, the fluctuations in SIDC data strongly influence solar
magnetism and
TSI variability. In this context, we suggest that these properties of SN
could be responsible
for the dynamics that drive the properties of the $q$--Triplet.}

Since the deviation of the three-index set ({\it q}$_{stat}$,{\it
q}$_{sen}$,{\it q}$_{rel}$) from unity is a measure of the departure from
thermodynamic
equilibrium, we also calculated the $q$'s indices for the present dataset,
in different timescales. For example, for $\tau$=1 day we found that the
three solar activity indices, SN, MF and TSI, may be essentially described
by the $q$--Triplet set ({\it q}$_{stat}$,{\it q}$_{sen}$,{\it q}$_{rel}$)
= (1.31$\pm$0.07,$-$0.71$\pm$0.10,1), (1.21$\pm$0.06,$-$0.44$\pm$0.07,1)
and \textcolor{black}{(1.54$\pm$0.03,$-$0.52$\pm$0.10,1)} respectively.
\textcolor{black}{Identical values of
the $q$--Triplet (within the uncertainties) are obtained
when we take into consideration the same dataset across the whole solar
cycle 23.
In addition, the multifractal spectra $f(\alpha)$ of the three datasets
observed across the whole solar cycle 23 are very similar to those
observed during the increasing phase of solar cycle 23. This result
suggests that solar magnetic activity tends to be in a quasi-stationary
and metastable
state for a same solar index. Nevertheless, when we take into account the
three solar indices,
measured simultaneously in the same period of the
solar cycle, we find that solar magnetic activity is non-stationary. Such
a comparison, between the whole cycle and its increasing phase, points in
fact for a relation between the timescale $\tau$ and the multifractal
spectrum asymmetry,
in the sense that the larger the timescale required to reach the
equilibrium state
is, the more the multifractal spectrum is asymmetric (see Figures
\ref{fig1} and \ref{fig2}).
In other words, there is a clear relation between indices $q_{stat}$ and
$q_{sen}$. In short, one sees
that the dual relation \cite{tsallis2004,queiros2007}
\begin{equation}
\label{dual}
q_{stat}+|q_{sen}|=2,
\end{equation}
is approximately satisfied (within the uncertainties) in the scenario of
solar magnetic activity. This is the first time, to our knowledge, that a
possible relationship between the $q_{stat}$ and $q_{sen}$ is obtained in
the study of solar phenomena.} \textcolor{black}{This duality reveals
that the subsystems here analyzed (represented by the three solar indices
Sunspot Number, Magnetic Field and Solar Total Irradiance)
are \textit{not asymptotically scale-invariant}, and that, consequently,
they are  \textit{strongly (or globally) correlated \cite{tsallis2005}}.
These facts strongly suggest that important
correlations exist between the random variables
involved in the relevant physical process controlling the solar 
activity, which still eludes us.
Finally, it is important to underline that these results are in clear
agreement with the $q$--CLT \cite{umarov2008,tsallis2009}.  Indeed, 
according this theorem, the $q$--Triplet seems to be related with 
duality relations.}

\acknowledgments
Research activities at the Stellar Board of the Federal University of Rio
Grande do Norte are supported by continuous grants from CNPq and FAPERN
brazilian agencies. D. B. de Freitas acknowledges a Ph.D. fellowship of
the CNPq brazilian agency. \textcolor{black}{The Sunspot data were made
available by Sunspot Index Data Center (SIDC). NSO/Kitt Peak data used
here are produced cooperatively by NSF/NOAO, NASA/GSFC, and NOAA/SEL. The
VIRGO TSI data are available from the SOHO VIRGO web site.}

\end{document}